\documentclass[a4paper,11pt]{article}
\usepackage{pos}

\usepackage{url}
\usepackage{colortbl}
\usepackage{booktabs}
\usepackage{MnSymbol}
\usepackage{bbold}

\usepackage{natbib}
\usepackage{adjustbox}
\usepackage{subfig}

\usepackage{lineno}
\definecolor{darkred}{rgb}{0.55, 0.0, 0.0}
\definecolor{crimsonglory}{rgb}{0.75, 0.0, 0.2}
\definecolor{sapphire}{rgb}{0.03, 0.15, 0.4}
\definecolor{ao}{rgb}{0.0, 0.5, 0.0}
\definecolor{huntergreen}{rgb}{0.21, 0.37, 0.23}
\definecolor{goldenyellow}{rgb}{0.99, 0.76, 0.0}

\definecolor{cernblue}{cmyk}{1.00,0.75,0,0}
\definecolor{cernblue80}{cmyk}{0.84,0.64,0,0}
\definecolor{cernblue60}{cmyk}{0.64,0.45,0,0}
\definecolor{cernblue40}{cmyk}{0.44,0.27,0,0}
\definecolor{cernblue20}{cmyk}{0.22,0.13,0,0}

\definecolor{cernblack}{cmyk}{0.91,0.79,0.62,0.97}
\definecolor{cernblack80}{cmyk}{0.69,0.60,0.56,0.66}
\definecolor{cernblack60}{cmyk}{0.56,0.45,0.45,0.33}
\definecolor{cernblack40}{cmyk}{0.41,0.32,0.32,0.11}
\definecolor{cernblack20}{cmyk}{0.23,0.17,0.18,0.01}

\title{Translating topological benefits in very cold lattice simulations}
\author[a,b]{Mattia Bruno}
\author[c]{Marco C\`e}
\author*[d]{Anthony Francis}
\author[e]{Jeremy R. Green}
\author[f]{Max Hansen}
\author[g]{Savvas Zafeiropoulos}

\newcommand\mib{Dipartimento di Fisica, Universit\'a di Milano-Bicocca, Piazza della Scienza 3, I-20126 Milano, Italy}
\newcommand\infn{INFN, Sezione di Milano-Bicocca, Piazza della Scienza 3, I-20126 Milano, Italy}

\affiliation[a]{\mib}
\affiliation[b]{\infn}

\affiliation[c]{Albert Einstein Center for Fundamental Physics (AEC) and Institut f\"ur Theoretische Physik, Universit\"at Bern, Sidlerstrasse 5, 3012 Bern, Switzerland}

\affiliation[d]{Institute of Physics, National Yang Ming Chiao Tung University, 30010 Hsinchu, Taiwan}

\affiliation[e]{Deutsches Elektronen-Synchrotron DESY, Platanenallee 6, 15738 Zeuthen, Germany}

\affiliation[f]{Higgs Centre for Theoretical Physics, School of Physics and Astronomy, The University of Edinburgh, Edinburgh EH9 3FD, UK}

\affiliation[g]{Aix Marseille Univ, Universit\'e de Toulon, CNRS, CPT, Marseille, France.}

\abstract{Master-field simulations offer an approach to lattice QCD in which calculations are performed on a small number of large-volume gauge-field configurations. The latter is advantageous for simulations in which the global topological charge is frozen due to a very fine lattice spacing, as the effect of this on observables is suppressed by the spacetime volume. Here we make use of the recently developed Stabilised Wilson Fermions to investigate a variation of this approach in which only the temporal direction ($T$) is taken larger than in traditional calculations. As compared to a hyper-cubic lattice geometry, this has the advantage that finite-$L$ effects can be useful, e.g.~for multi-hadron observables, while compared to open boundary conditions, time-translation invariance is not lost.\\
\vspace{-1ex}

In this proof-of-concept contribution, we study the idea of using very cold (i.e.~long-$T$) lattices to topologically "defrost" observables at fine lattice spacing. We identify the scalar-scalar meson two-point correlation function as a useful probe and present first results from $N_f=3$ ensembles with time extents up to $T=2304$ and a lattice spacing of $a=0.055$\,fm.}

\FullConference{%
 The 39th International Symposium on Lattice Field Theory (Lattice2022),\\
 8-13 August, 2022 \\
 Bonn, Germany
}

\begin{document}
\hfill{DESY-22-205}
\maketitle

\section{Introduction and formalism}

Lattice QCD is entering a precision era and the results are impacting some of the most interesting areas of particle physics. With many observables demonstrating good control over finite-volume and quark-mass effects, discretisation effects are becoming more central as dominant sources of systematic uncertainty, see e.g. \cite{Borsanyi:2020mff,Chimirri:2022gzg}. Future progress will rely on increased control over the continuum limit, and expanding current parameter ranges towards finer lattice spacings is becoming an important goal.

With current algorithms, as the lattice spacing is decreased, the tunnelling probability between topological sectors drops. Consequently, the autocorrelation time between independent configurations increases without bound and topology freezes. A simulation with frozen topology, i.e.~at fixed topological charge $Q$, does not truly sample QCD; rather, it can be interpreted as inserting an unwanted power series in $Q$ into QCD correlation functions. Of course ensuring that the expectation value of $Q$ vanishes is, by itself, not sufficient to guarantee that freezing effects are under control. %An important counter-example in this respect is parity-odd correlators, considered in more detail in the following sections.

The work presented here is based on the observation that the effects of frozen topology are suppressed by at least one power of $1/V$, where $V=L^3T$ is the spacetime volume of the simulation~\cite{Brower:2003yx,Aoki:2007ka}. In this proof-of-concept contribution, we explore the idea of using lattices with a very large $V$, induced by a very long time direction, to mitigate these contaminations.

\subsection{Open boundary conditions}

The problem of the freezing of global topology can, in principle, be elegantly solved via open boundary conditions, see \cite{Luscher:2011kk,Luscher:2012av}. By replacing the usual (anti-)periodic boundary conditions in the time direction with Neumann conditions, it becomes possible for topological objects to diffuse in and out of the lattice through the boundaries. Topology continues to evolve in Monte Carlo (MC) time, also at fine lattice spacings.

However, the open boundaries entail the loss of translational invariance in the time direction. The consequences of this must be considered on an observable-specific basis. For example, for spectral quantities such as particle masses, the boundaries are less important since the energies in the spectral decomposition will be unaffected and the only consequence of operators approaching the boundary is a modification to the overlap factors. With other observables the effect is of direct relevance. For example for the hadronic-vacuum-polarization contribution to the muon's magnetic moment, defined via a vector-vector correlator, both vector insertions must be sufficiently far from the boundary to ensure one is estimating a vacuum expectation value. In the following we show an example of an observable that shows visible contamination due to this effect.

\subsection{An alternative concept: The long-$T$ approach}

Among other ideas to address the topology freezing problem at acceptable computational cost, one new path is to note that, in QCD, spatially distant regions largely fluctuate independently of each other. In ref.~\cite{Luscher:2017cjh} it was shown how this can change our perspective of building expectation values $\langle ... \rangle$ via averages over MC time histories ($\bar{...}$), into one in which we understand the same process as a translational averaging over locally decorrelated spacetime regions, denoted $\llangle ... \rrangle$:
\begin{equation}
\llangle \mathcal{O}(x) \rrangle = \frac{1}{V} \sum_{z}\mathcal{O}(x+z),\qquad \qquad \langle \mathcal{O}(x) \rangle = \llangle \mathcal{O}(x) \rrangle + O(V^{-1/2}) \,.
\end{equation}
Just as one can estimate the error from MC samples, taking into account autocorrelations, it is also possible to estimate the error from translated samples, accounting for spatial correlations. This implies that in the extreme regime of very large volumes, i.e.~where the four-volume is significantly larger than in typical present-day calculations, it is sufficient to have just a single gauge field configuration to estimate both the central value and the uncertainty of any observable. This  "master-field" (MF) approach makes effective use of the stochastic locality property of QCD.\footnote{Progress in generating master-fields has been reported recently, also at this conference \cite{Fritzsch:2021klm,Fritzsch:2022lattice}.}

Aside from a new way of statistically evaluating observables and their uncertainties, this also suggests a way to side-step the topology freezing problem. As already mentioned in the introduction, the effect of freezing scales as $1/V$ as for sufficiently large $V$. Statistical uncertainties, by contrast, are only suppressed by $1/\sqrt{V}$ for a fixed number of gauge fields, and are thus guaranteed to dominate over freezing effects, making the latter irrelevant, as the spacetime volume increases.

From the point of view of topological suppression, the MF regime is reached through any increase in the spacetime volume. The focus of this work is to take only one direction large. To state the idea compactly; if $L_{\rm trad}$ and $T_{\rm trad}$ denote traditional spatial and temporal extents, respectively, we take
\begin{equation}
L=L_{\rm trad},~ T\gg T_{\rm trad} ~\longrightarrow~ \textrm{long-}T\textrm{ approach}\,.
\end{equation}

An advantage of this version of the MF in the context of hadron spectroscopy is that the spectrum set by the spatial volume remains sparse. This implies, e.g., that the established and advanced methods of the finite-volume formalism can be applied as usual.\footnote{In contrast, infinite-volume based methods, including those that make use of position-space correlators, are attractive in the MF approach and being explored in refs.~\cite{Ce:2021akh,Ce:2022lattice}.} The smaller spatial volume is also useful for methods based on identifying the low modes of a three-dimensional operator, such as distillation. Furthermore, the approach keeps translational invariance, which allows one to exploit the volume more fully. Starting to explore the long-$T$ approach, our goals here are to generate the first long-$T$ lattices in the regime of slowed topological sampling and to identify sensitive probes to establish their benefits.

\begin{table}[t!]
 \centering
 \begin{tabular}{lccrcccrcc}
 %\begin{tabular}{cc>{\columncolor{yellow!20}\color{black}}rrc>{\columncolor{yellow!20}\color{black}}cr}
 \toprule
 Label & {$a$[fm]}, {$m_\pi$[MeV]} & {$L/a$} & {$T/a$}& {$N_{\rm{cfg}}$} & BC's & $\bar{Q}$ & $V_{\rm{rel}}$ & {$\tau_Q$} & {$\tau_E$} \\\hline
 P(96) & 0.055, 418 & 48 & 96 & 488 & P & 1.3 & 1 & 11(4) & 5(2) \\
 P(384) & & & 384 & 101 & P & 3.0 & 4 & 3(1) & 4(2)\\
 P(1152) & & & 1152 & 94 & P & -8 & 12 & 4(2) & 3(1) \\ \hline
 P(2304)$_1$ & & & 2304 & 38 & P & -50 & 24 & 2(1) & 2(1)\\
 P(2304)$_2$ & & & 2304 & 36 & P & -12 & 24 & 2(1) & 3(1)\\ \hline
 O(96) & & & 96 & 495 & \textcolor{black}{O} & -1.0* & 1 & 7(3)* & 3(1)\\
 \toprule
 \end{tabular}
 \caption{Ensemble parameters. We define $V_{\rm rel}=V/V_{96}$; $\tau_Q$ and $\tau_E$ denote integrated autocorrelation times of $Q$ and $E$. "*"= with open boundary conditions $\bar Q$ and $\tau_Q$ are not well defined, see text. }
 \label{tab:ensembles}
\end{table}

\begin{figure}
\centering
\includegraphics[width=0.475\textwidth]{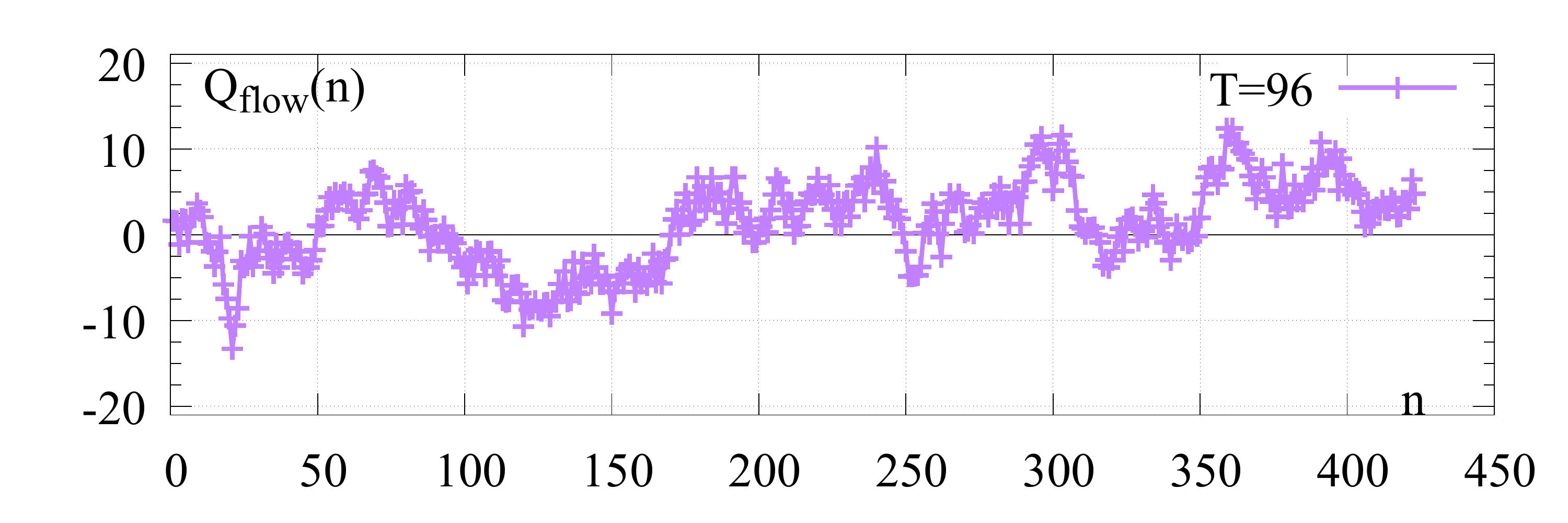}
\includegraphics[width=0.475\textwidth]{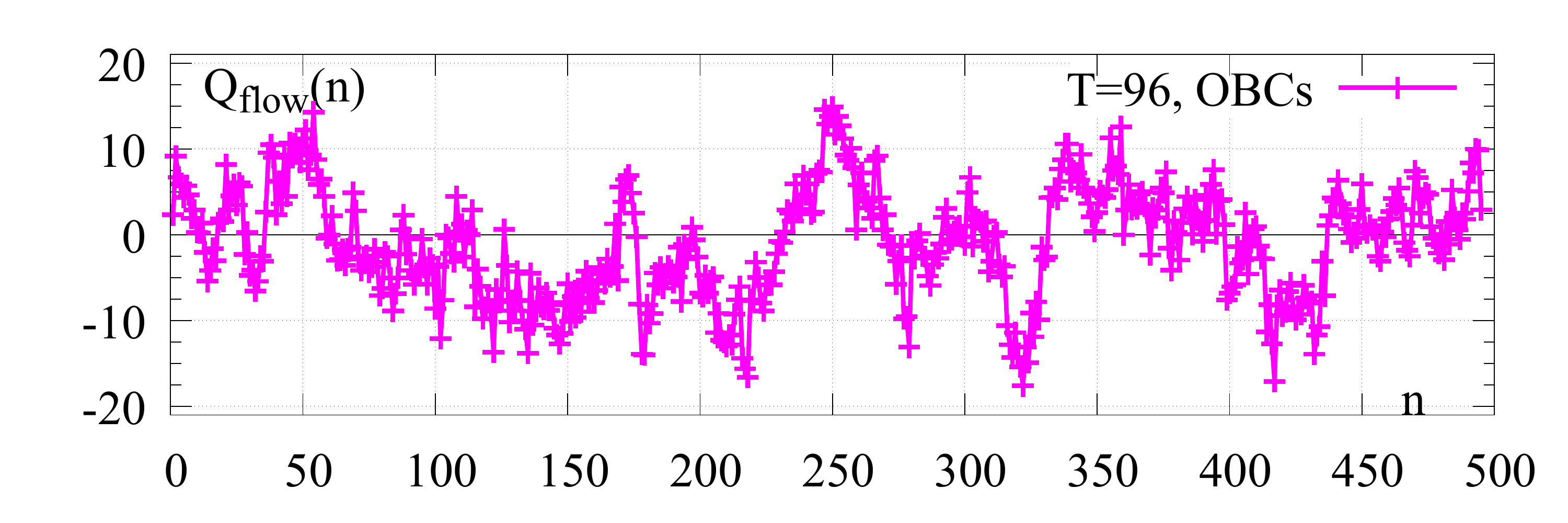}
\includegraphics[width=0.475\textwidth]{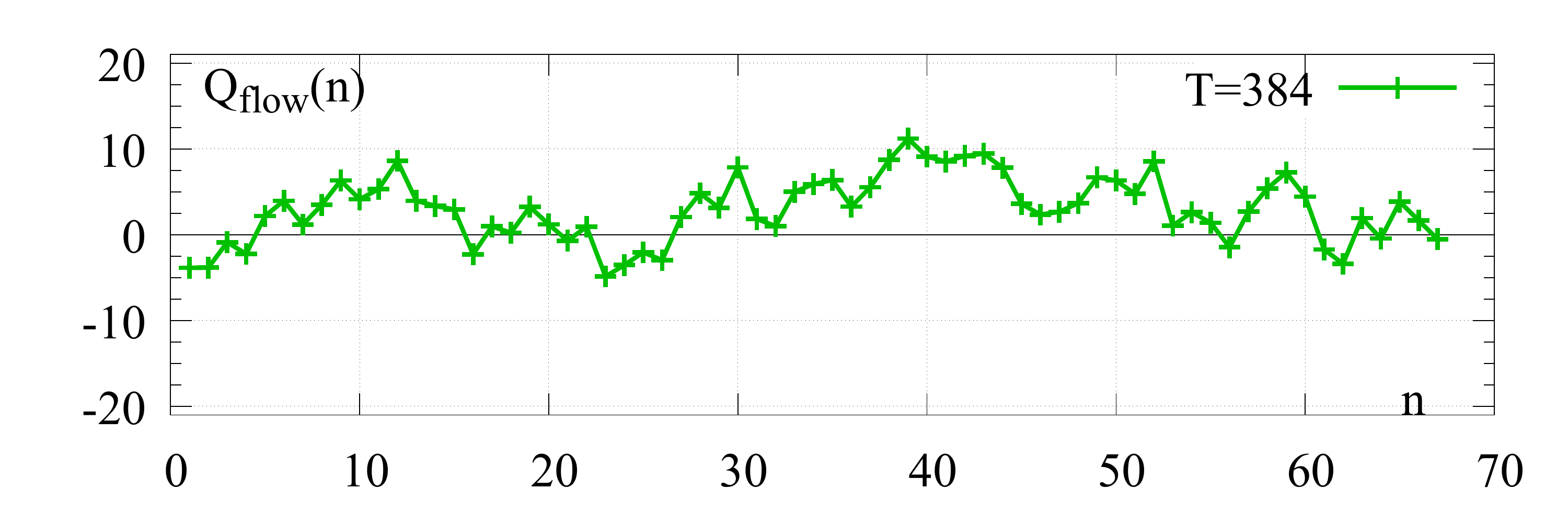}
\includegraphics[width=0.475\textwidth]{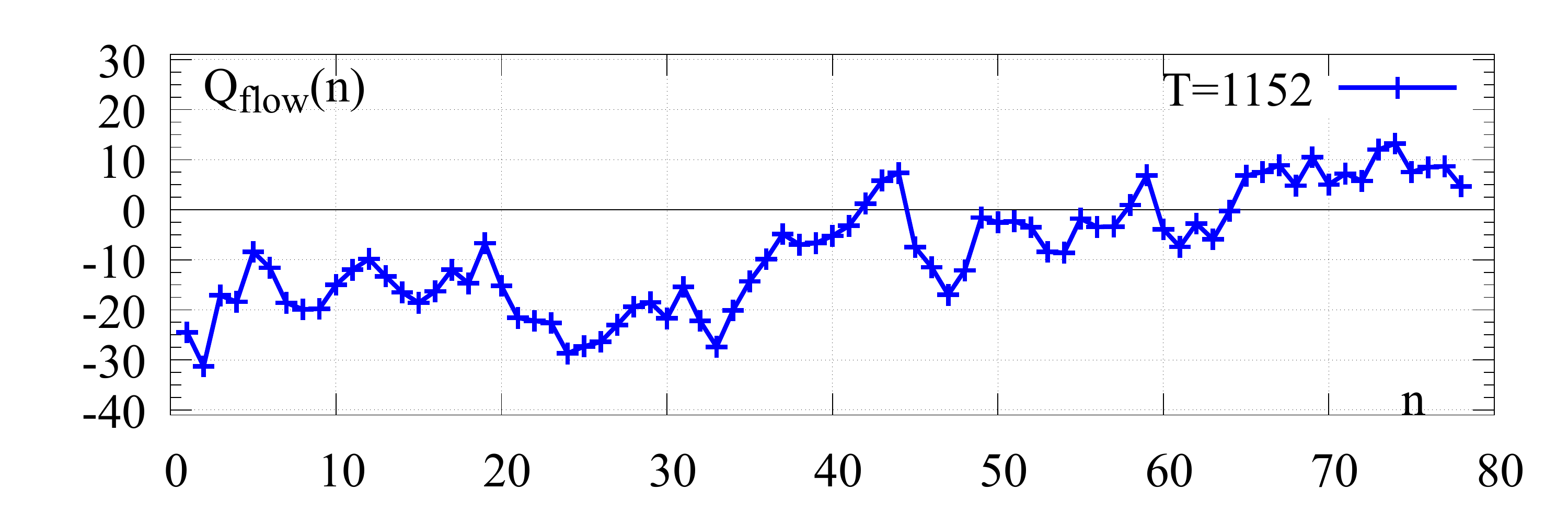}
\includegraphics[width=0.475\textwidth]{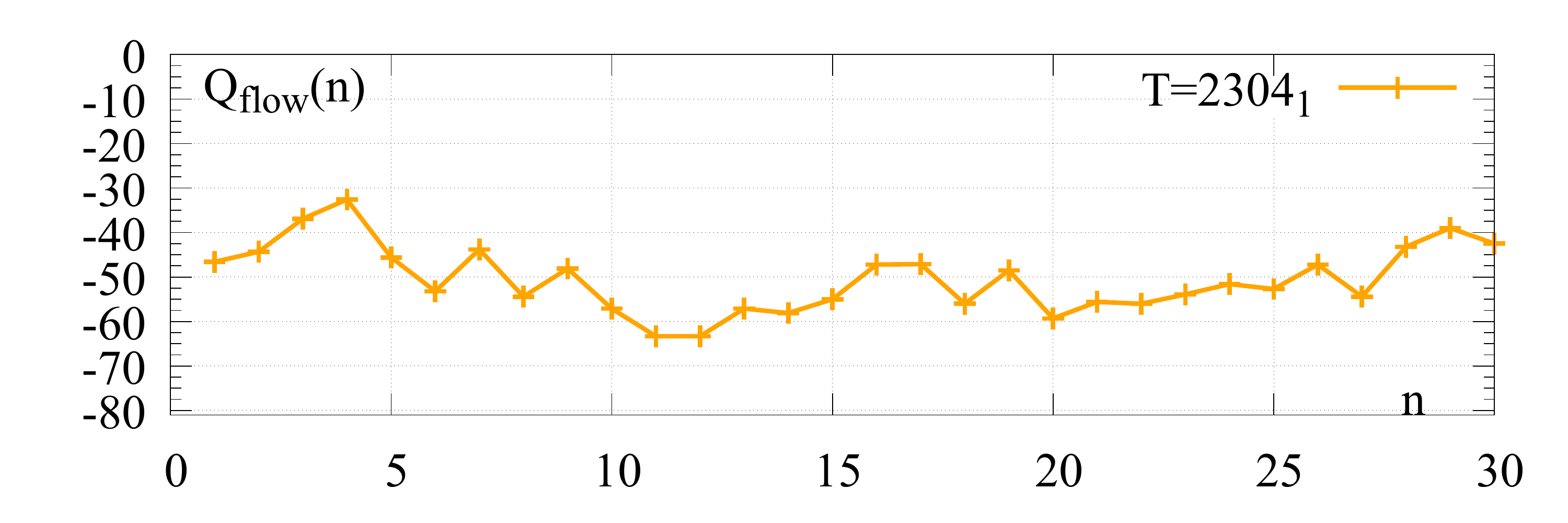}
\includegraphics[width=0.475\textwidth]{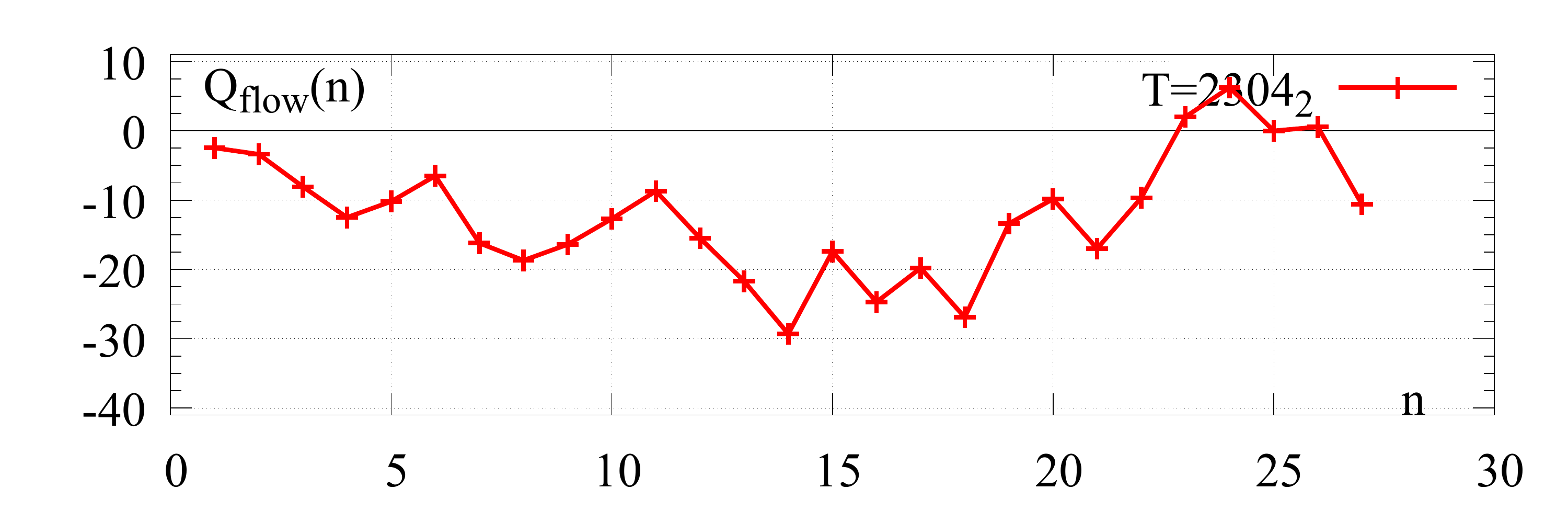}
\vspace{-1ex}
\caption{Topological charge MC histories for all ensembles. The values for $Q$ with open boundaries are given here for reference only. Note the distinct topological sectors for the two separate $T/a=2304$ strings.}
\label{fig:qcharge}
\end{figure}

\section{Numerical setup}
%\section{Numerically setting up long-$T$ ensembles}

To test this new approach the first aim is to generate dynamical configurations in a regime where topology freezing is a problem.
To this end, a setup with $N_f=2+1$ dynamical fermion flavors at the single lattice spacing of $a=0.055$~fm is chosen. It is set via the gradient flow scale and converted to physical units via $\sqrt{8t_0}=0.414(5)\textrm{fm}$~\cite{Bruno:2016plf}. Furthermore, the stabilised Wilson fermion framework (SWF) \cite{Francis:2019muy} is used as it cures a large volume pathology of more common Wilson-Clover actions. The exponentiated Clover coefficient is non-perturbatively set according to \cite{Francis:2019muy,Francis:2022hyr}; see also \cite{Cuteri:2022gmg} for more details on the setup.
Previous studies (for the SWF case see for instance \cite{Cuteri:2022gmg}) show that at this lattice spacing topological sampling slows down to a point where a switch to open boundaries becomes beneficial.
The masses of the three quark flavors are set equal and are fixed to $m_\pi=m_K=418$~MeV.\footnote{Note, this is slightly away from the so-called flavor symmetric point where $m_\pi=m_K=412$~MeV\cite{Bruno:2014jqa}.}

%\subsection{Towards long-$T$ ensembles}

Throughout the spatial volume is fixed to $L/a=48$, corresponding to $m_\pi L=5.6$, while the time extent is increased 
from 96 to 2304 points. At $T/a=96$ two different ensembles are generated, one with periodic and one with open boundary conditions.
For all long-$T$ ensembles the boundary conditions are periodic. An up-folding strategy is used where the smaller time extent configurations seed the larger ones.
The expected slow evolution of topology in these simulations offers a novel way to test the volume suppression effect:
We choose to generate two separate strings at $T/a=2304$ seeded from smaller volume configurations in distinctly different topological sectors. Moving away from these sectors only very slowly in the MC evolution, a situation is engineered where observables carry different contamination and the volume suppression can be tested explicitly.

The ensemble parameters are gathered in Tab.~\ref{tab:ensembles}, in particular note the inclusion of the relative volume factor $V_{\rm rel}=V/V_{96}$. The reference lattice volume, $L/a=48$, $T/a=96$, was chosen to reflect a common, traditional lattice setup.

%begin{figure}
%\centering
%\includegraphics[width=0.39\textwidth]{./pics/plaq.pdf}
%\includegraphics[width=0.39\textwidth]{./pics/yact.pdf}
%\includegraphics[width=0.39\textwidth]{./pics/rwf.pdf}
%\caption{ $log(x)$ axis for legibility, opening spread is optical effect}
%\label{fig:plaqs}
%\end{figure}

%To illustrate the generation process the MC histories of the average plaquette $P_n$, $t^2\langle E \rangle_n$ and the reweighting factor $RWF_n$ are shown in Fig.~\ref{fig:plaqs}. The $x$-axes are given on logscale to increase visibility of the large $T$ ensembles that have generally shorter MC chains. We do not observe any immediate oddities or thermalisation effects.

\begin{figure}[t!]
\begin{center}
\includegraphics[width=0.32\textwidth]{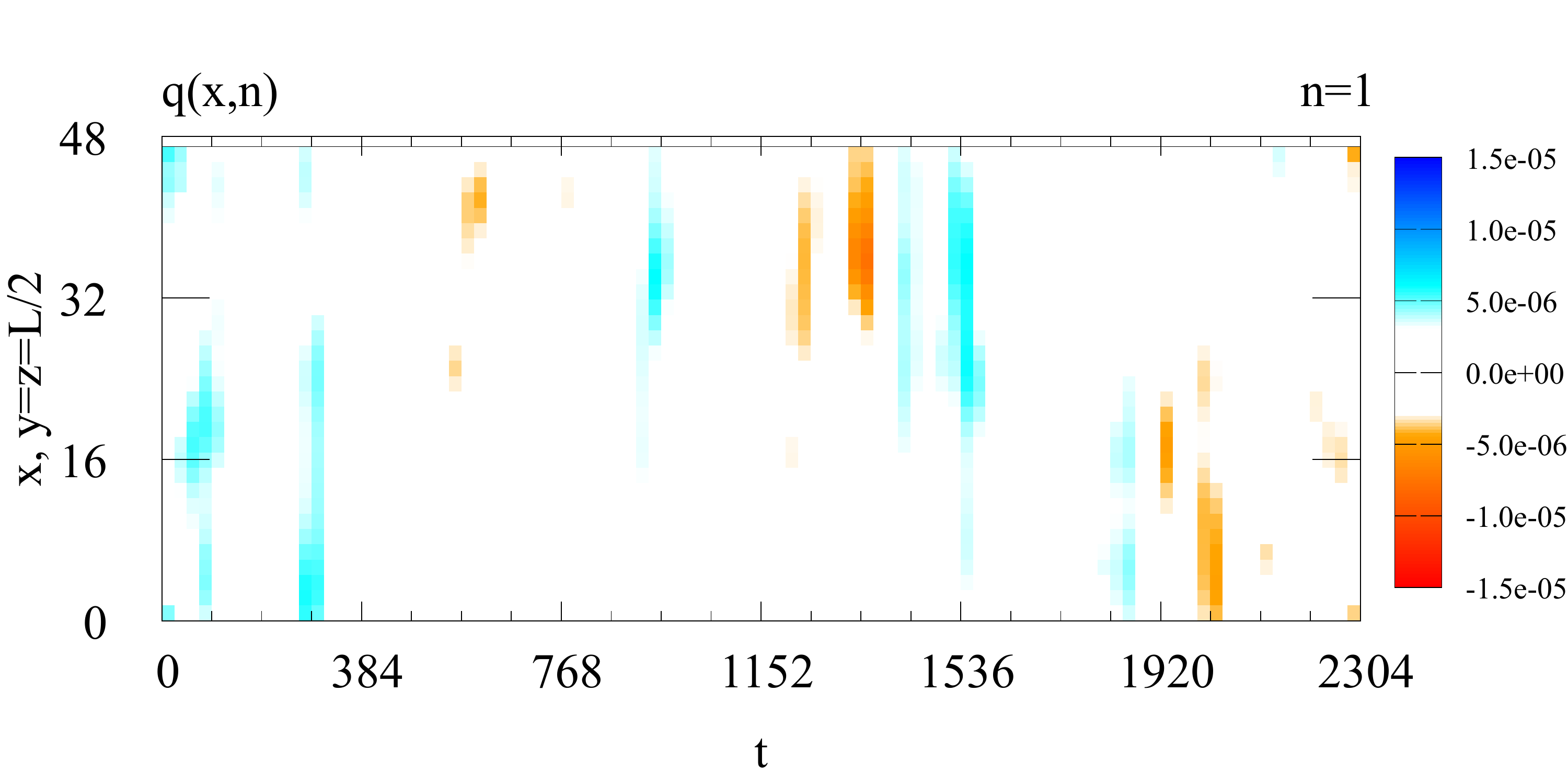}
\includegraphics[width=0.32\textwidth]{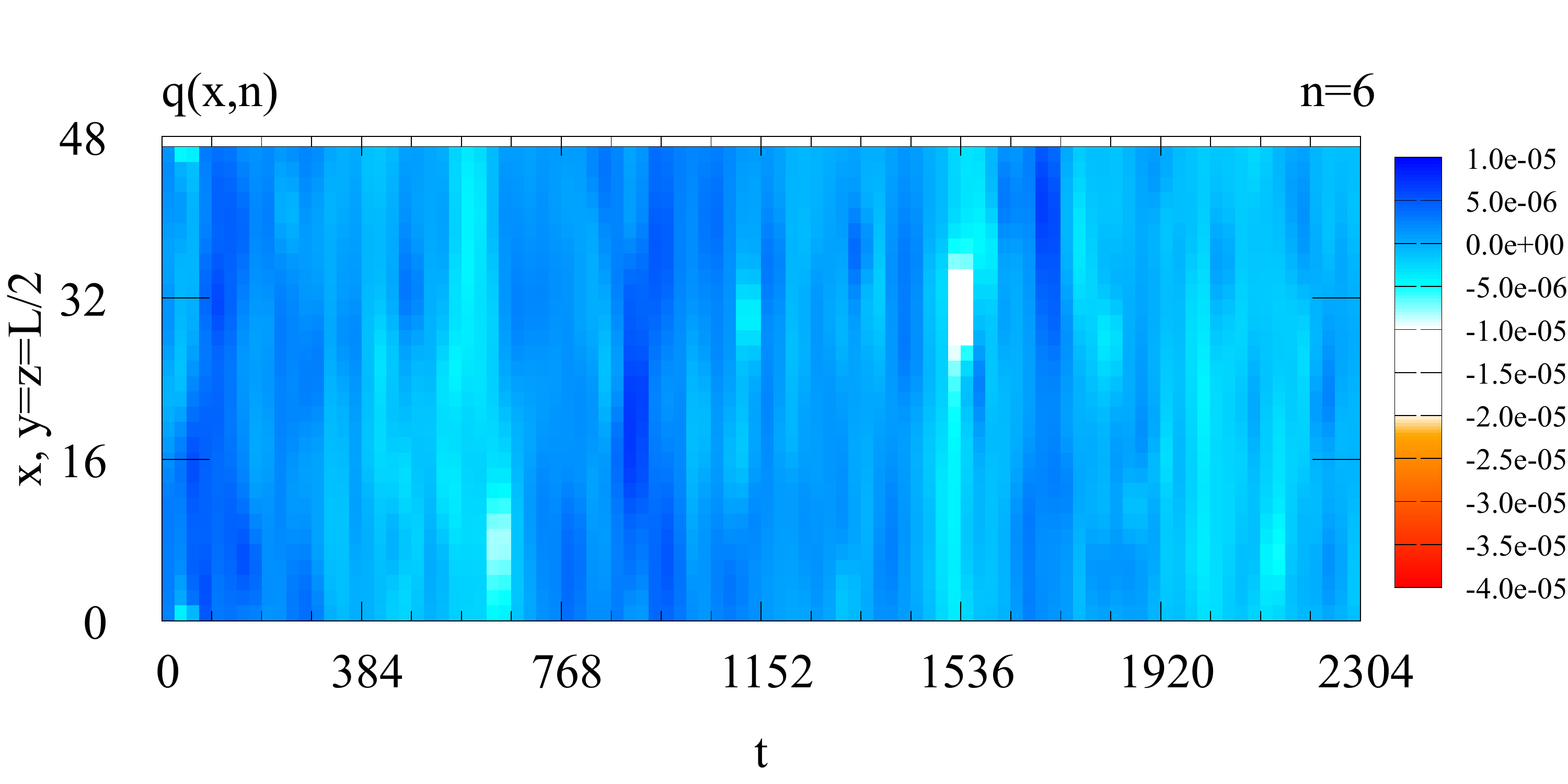}%\\
%\vspace{-3ex}
\includegraphics[width=0.32\textwidth]{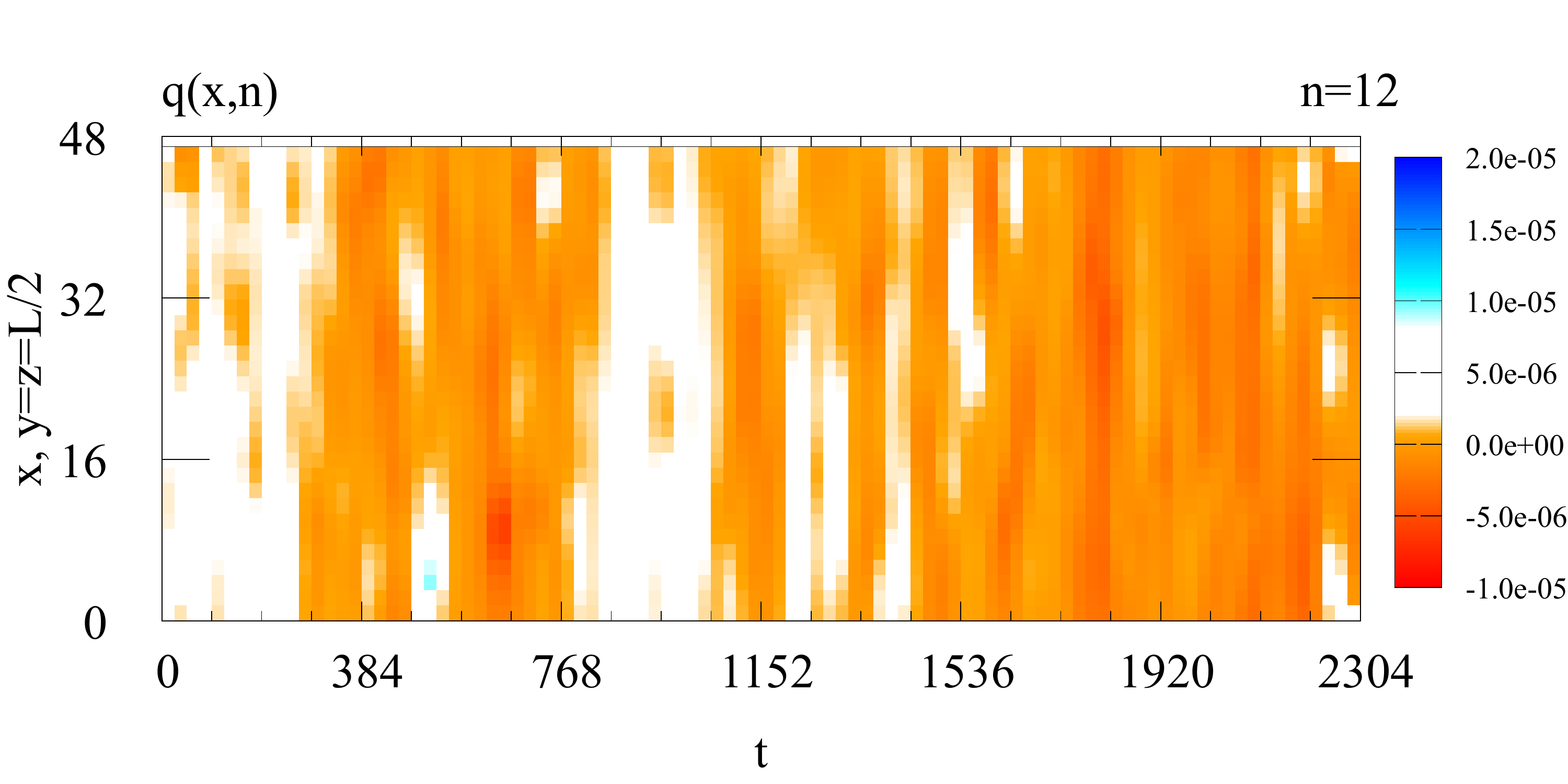}
\includegraphics[width=0.32\textwidth]{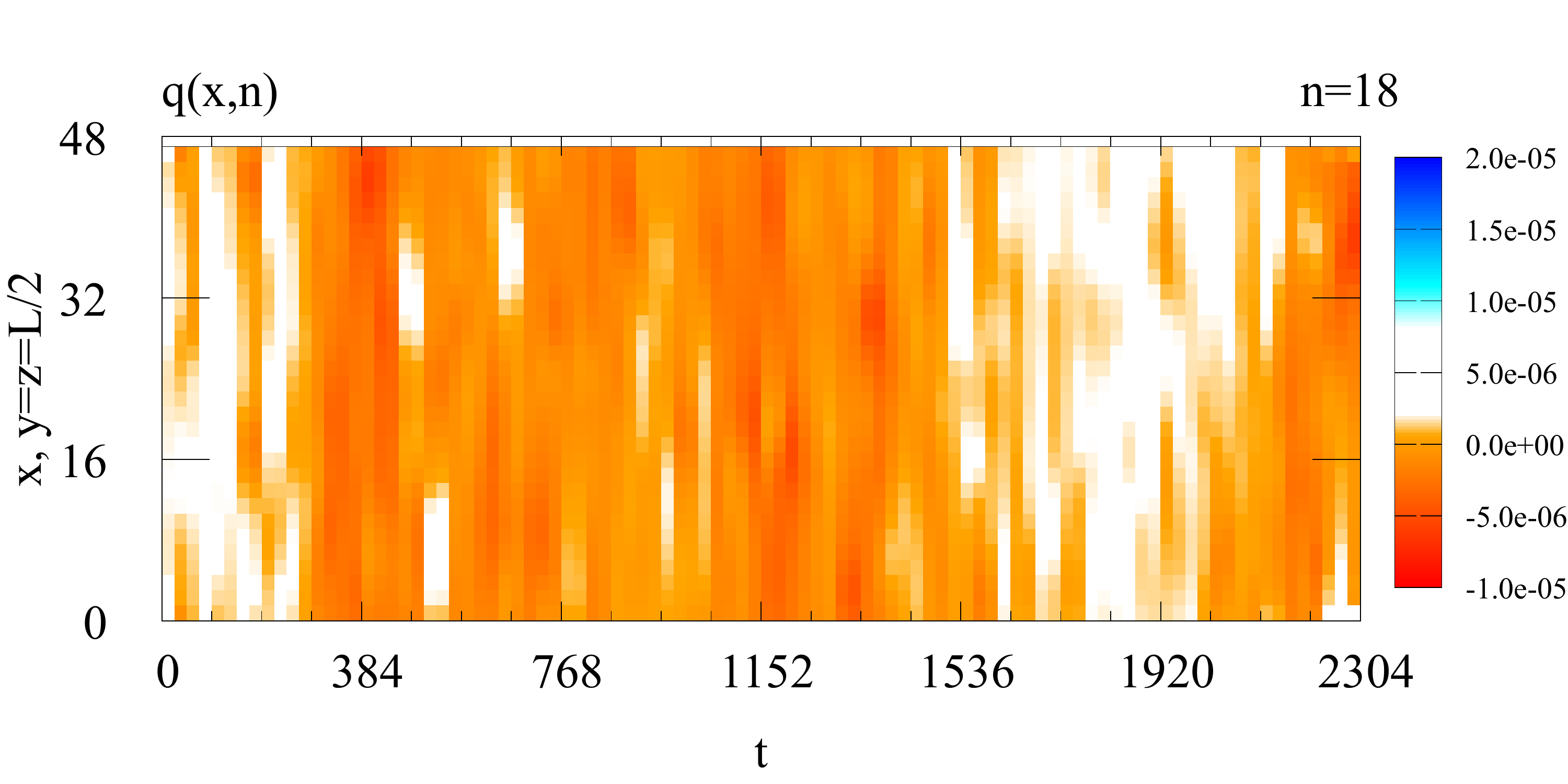}%\\
%\vspace{-3ex}
\includegraphics[width=0.32\textwidth]{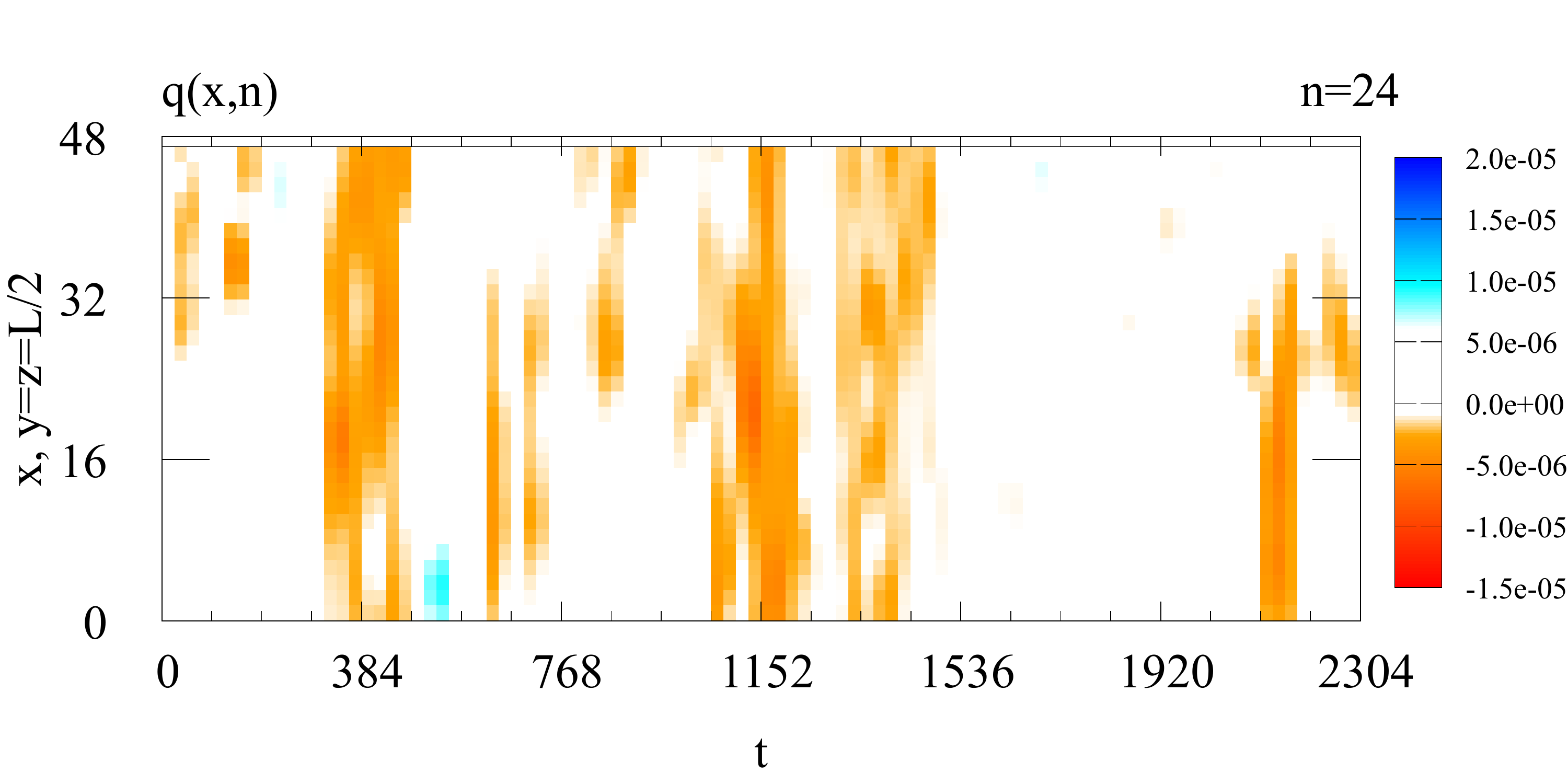}
\includegraphics[width=0.32\textwidth]{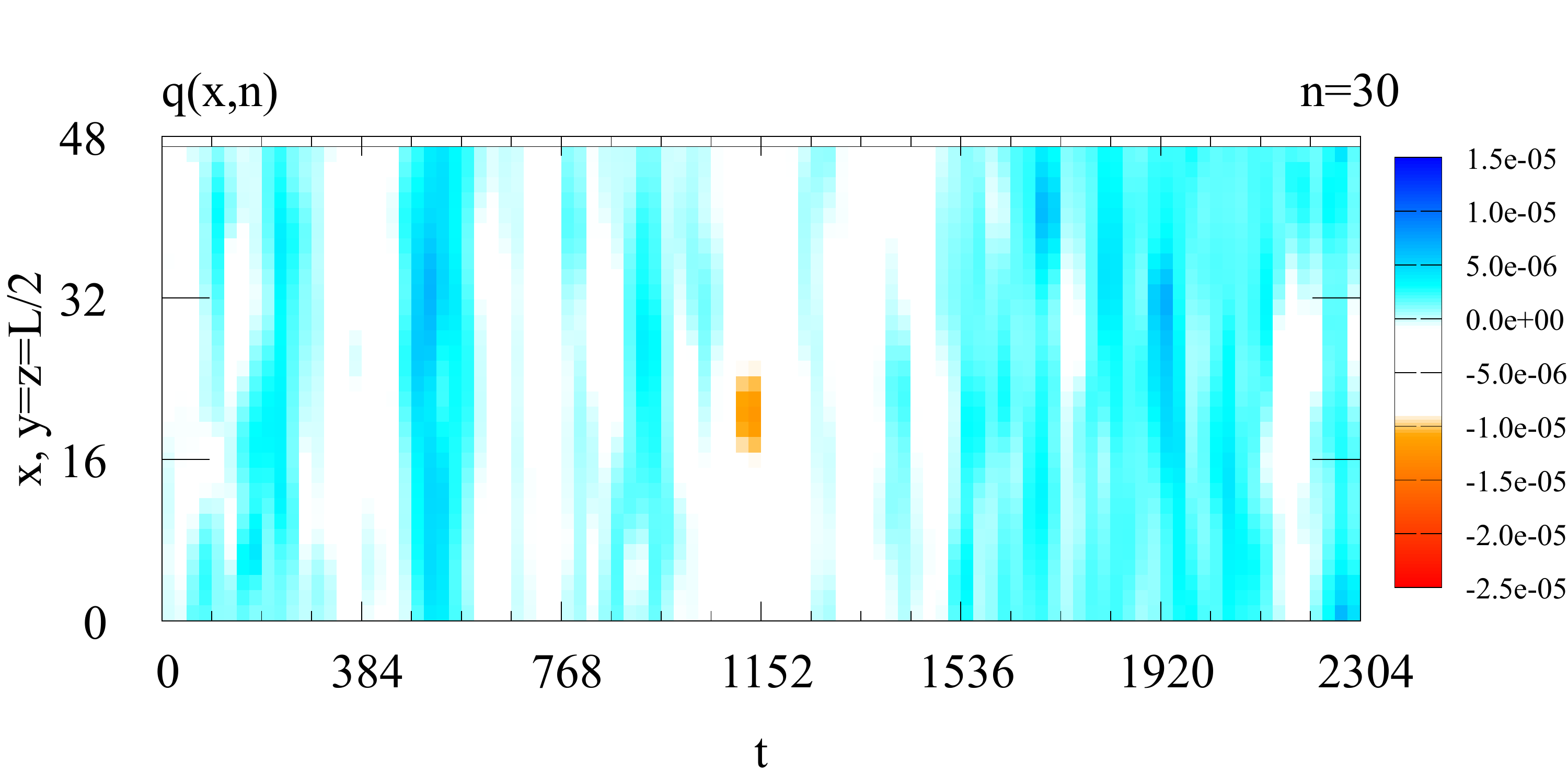}
\end{center}
\vspace{-3ex}
\caption{Quantity $q(x)$ of the $T/a=2304$ lattice with $\bar{Q}=-12\sim \sum q(x)$ in space-time slice $x,t$ where $y=z=L/2$. From left to right and top to bottom the panels show the configurations $n=1,6,12,18,24,30$. }
\label{fig:thermal}
\end{figure}

\section{Observations during generation and thermalisation}

In the preceding discussion the topological charge is identified as a key observable. Here, we follow the gradient flow based definition \cite{Luscher:2010iy}, evaluated at positive flow time $t_{\rm flow}=1.3t_0=9.6\,a^2$:
\begin{equation}
Q = \sum_{x\in V} q(x)~~,~~~q(x) = - \frac{1}{32\pi^2} \epsilon_{\mu\nu\rho\sigma} {\rm Tr}[ F_{\mu\nu}(x)F_{\rho\sigma}(x)]~.
\end{equation}

The MC averages $\bar{Q}$ and MC histories of this observable are given in Tab.\ref{tab:ensembles} and Fig.~\ref{fig:qcharge} for all six ensembles, respectively.
Note, that the topological charge is not cleanly defined for open boundary conditions; for reference purposes we show here the measurement of $\bar{Q}$ computed on a volume that allows instantons to flow in and out. As such, it is not quantized (integer) in the continuum limit.

Throughout we observe a slow evolution of the topological charge with MC time. Given the short MC chain lengths we forgo quoting an error for the average charge given in Tab.~\ref{tab:ensembles} at this point. We observe that the two $T/a=2304$ ensembles indeed remain in their distinct topological sectors, one string exhibits an average topological charge of $\bar{Q}=-50$ and the other one of $\bar{Q}=-12$ with similar chain lengths. In particular the chains stay apart from each other throughout their evolution history.
To visualise the evolution of local topological fluctuations, we show $q(x)$ on the ensemble P(2304)$_2$ for the space-time slice $x,t$ where $y=z=L/2$ for a number of points in MC time in Fig.~\ref{fig:thermal}.
This preliminary check shows significant evolution away from the initial, seed configuration.
The integrated autocorrelation times $\tau_Q$ and $\tau_E$, of the topological charge and the energy density at $t_0$, respectively, are given in Tab.~\ref{tab:ensembles}. Keep in mind that these estimates are based from relatively short MC chains. The autocorrelation times measured compare well with the distance in MC time between two visually distinct $q(x)$ fields.

Effective thermalisation is a key question in large volume or MF-type simulations. At this stage we did not observe signs of thermalisation effects. However, we must leave a quantitative study for the future.

\begin{figure}
\centering
\includegraphics[width=0.49\textwidth]{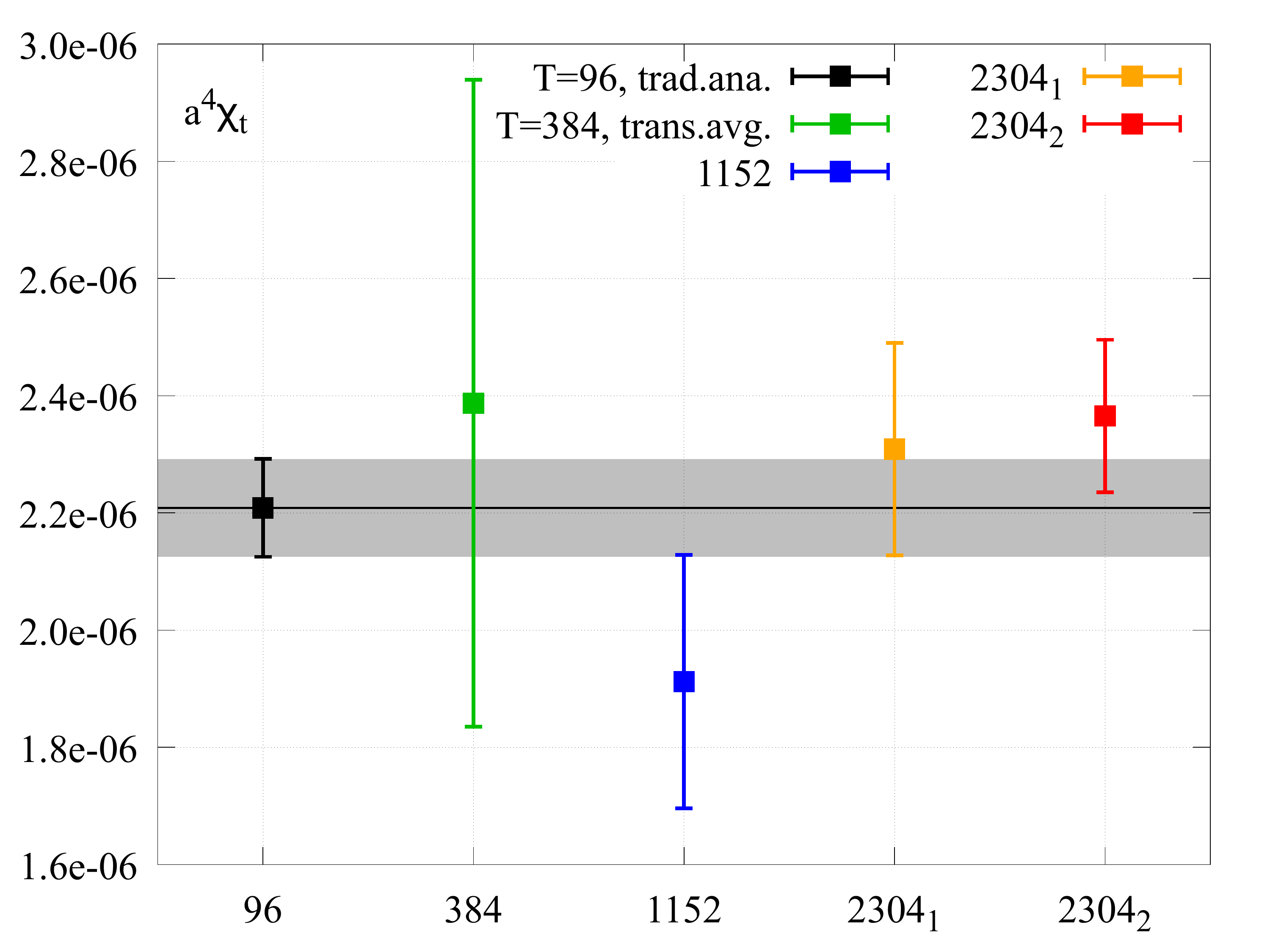}
\includegraphics[width=0.49\textwidth]{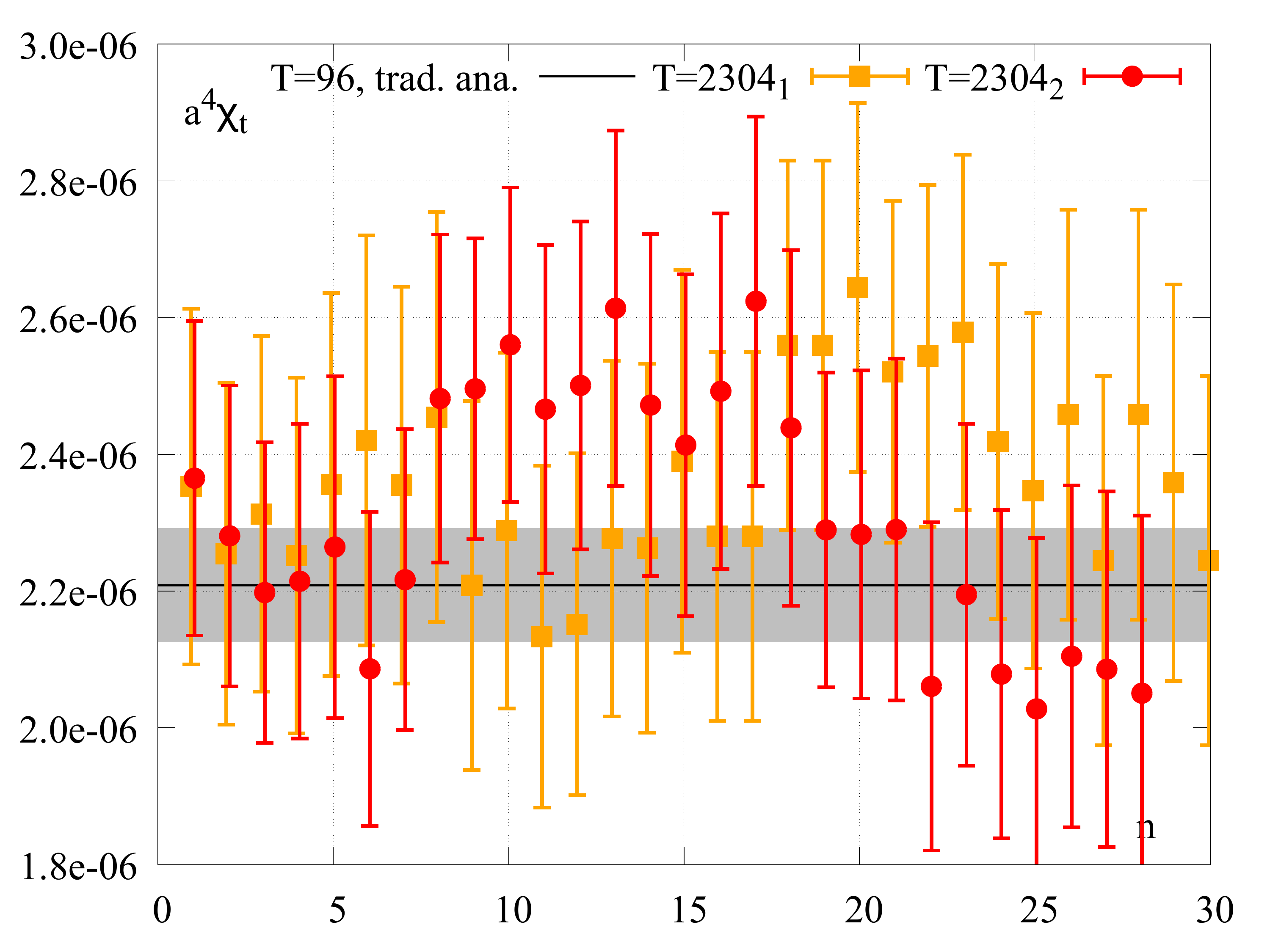}
\caption{Topological susceptibility. Left: Average values derived using a traditional error analysis on P(96) and translational averages on all other ensembles. Right: MC history of the average value (config-by-config) and the translational error on the two $T/a=2304$ ensembles. The results have been slightly horizontally shifted relative to each other for legibility.}
\label{fig:susc}
\end{figure}

%\section{Volume suppression of topological effects}
\section{Translational averages for the topological susceptibility }

Next, we study whether translational averaging is already effective in the $V_{\rm rel}$ reached here. To this end we follow the procedure outlined in \cite{Luscher:2017cjh}. 
Summarising it here, note the topological susceptibility is defined as:
\begin{equation}
\chi_t := \frac{\langle Q^2\rangle}{V}= \sum_{y} \langle q(y) q(0) \rangle ~.
\end{equation}
The sum over $y$ may be split into two parts on large lattices:
\begin{equation}
\chi_t = \sum_{|y|\leq R} \llangle q(y) q(0) \rrangle~ +~ \sum_{|y|> R} \langle q(y) q(0) \rangle ~+~ \mathcal{O}(V^{-1/2})~~,
\end{equation}
whereby the second term decays quickly at large $R$ and the translational average $\llangle ... \rrangle$ requires the addition of the $V^{-1/2}$-term. 
In practice this formulation enables the determination of $\chi_t$ by reading it off from the plateau value at large $R$.
The preliminary results of applying this procedure to all ensembles except for the smallest, $T/a=96$, are shown in Fig.~\ref{fig:susc} (left). 
For the latter an effective translational averaging cannot be expected and the errors are estimated using a bootstrap analysis in a more traditional analysis instead. Throughout we take care that all measurements have the same statistics when assuming that the error follows the naive relations with the volume and number of configurations. 
Comparing the results, for the ensembles P(384) and P(1152) we observe larger relative errors and some deviation. For the two P(2304)$_{1,2}$ ensembles, however, we observe similar errors and furthermore the values from both strings agree. %Keep in mind that the results are derived from strings that are in topological sectors differing on average by a factor four.
This is further illustrated in Fig.~\ref{fig:susc} (right) where the MC histories for these two lattices are given. We obtain consistent results from analysing each individual configuration. Recall translational averages are used here. 
Taken together we conclude that the regime where translational averaging becomes effective is reached for the longest lattices and the time extent is long enough to suppress the topological contamination below the level of the error in these lattices.

\section{Translating topological benefits and defrosting hadronic observables}

One of the motivations of this work is to explore new ways for topologically safe simulations for spectroscopy applications. 
Here, we focus on calculating zero-momentum meson correlation functions:
\begin{equation}
 G_{\mathcal{O}_1\mathcal{O}_2}(t=t'-t_{\rm src})=\sum_x \langle \mathcal{O}_2(x,t') \mathcal{O}_1(x_{\rm src},t_{\rm src}) \rangle~,
\end{equation}
where $\mathcal{O_i}=\bar \psi \Gamma_i \psi$ and $\Gamma=\gamma_5,\mathbb{1}$. Only connected channels are considered in this study. The correlators are denoted $\mathcal{O}_1\mathcal{O}_2\,\hat{=}\,G_{\mathcal{O}_1\mathcal{O}_2}(t)$ in the following. 

On the ensembles with open boundary conditions $U(1)$ noise wall sources are used. The sources are either placed close to the boundary, $t_{\rm src}=1,T/a-1$, or in the central region, $t_{\rm src}/T=(1/4),~(3/4)$.

On the periodic ensembles a similar setup is used, however there are important modifications: 
The immediate benefit of the long-$T$ approach is that it preserves translation invariance. One by-product of having a highly elongated time direction is that multiple sources can be safely put on a single configuration simultaneously as a consequence of stochastic locality. The around-the-world effects become negligible but are replaced with next-neighbour-wall effects\footnote{With a different noise setup, the 
next-neighbour-wall effects could be cancelled stochastically. This is not done here.}.
As a consequence we place $N_{m}=T/\delta t_{m}$ of $U(1)$ noise wall sources per configuration per inversion and spread them with $\delta t_{m}=96a$ starting from a randomly varied $t_{\rm src}$. 
While for P(96) this is very similar to the open boundary case, for all other ensembles this leads to a new kind of source consisting of multiple noise walls a distance $\delta t_{m}$ apart from each other. 
For example, on the $T/a=2304$ ensembles there are 24 walls that are solved for simultaneously in one inversion. The periodic images of the correlators obtained from this inversion are summed as initial analysis step. As such this wall array (warray) estimator of the correlator reads:
\begin{equation}
G_{\mathcal{O}_1\mathcal{O}_2}^{\rm warray}(t)=\sum_{n=0}^{N_m-1} G(t+n\cdot\delta t_m).
\end{equation}
In principle different $\delta t_{m}$ can be chosen according to the needs of the spectral analysis and noise properties of the correlator, e.g. one may choose to use $\delta t_{m}=192a$ or more for the pion as one can follow the signal for long distances. 
In this contribution the aim is compare results on the different ensembles and setups. For this reason we choose the same number of time slices $T=\delta t_{m}=96a$ and the number of sources, noises and configurations used is fixed such that based on naive error and volume scaling we can expect the same level of statistical uncertainty for all correlator measurements. Note, the wall sources introduce an additional relative factor $(L/a)^3$.

A good hadronic probe should be sensitive to the topology. We identify meson correlators of scalar ($\Gamma=\mathbb{1}$, denoted $S$) and pseudoscalar ($\Gamma=\gamma_5$, denoted $P$) operators as ideal in this sense: 
The $SS$ correlator can be distorted by topological effects even though it is parity-even, similar in spirit to the arguments of \cite{Aoki:2007ka,Bali:2014pva}.
One may expect the leading effect to be proportional to the difference between $Q^2/V$ and $\chi_t$.
At the same time, the $PS$ correlator is parity-odd and vanishes in QCD, i.e.~it should be zero (stochastically) in a simulation that correctly samples topology,
while in the case of topological contamination the $PS$ obtains non-zero signal. The correlator then behaves as:
\begin{align}
G_{PS}(t) \sim A_{PS}\cdot \exp[ - {m_\pi} t] ~~~~
\textrm{and} ~~~~A_{PS}~\sim Q/V~~.
\end{align}
In the following this effect in the amplitude $A_{PS}$ will be checked explicitly.
Furthermore, inserting $Q^2$ into the $SS$ correlator at long distance creates or annihilates a pion, as such one may expect it to decay towards $m_\pi$ asymptotically. We leave this to a future study.
In both cases the $PP$ channel gives a clean observable to test these predictions against.
Note that these statements are independent of the choice of boundary conditions.

\begin{figure}
\begin{center}
\includegraphics[width=0.49\textwidth]{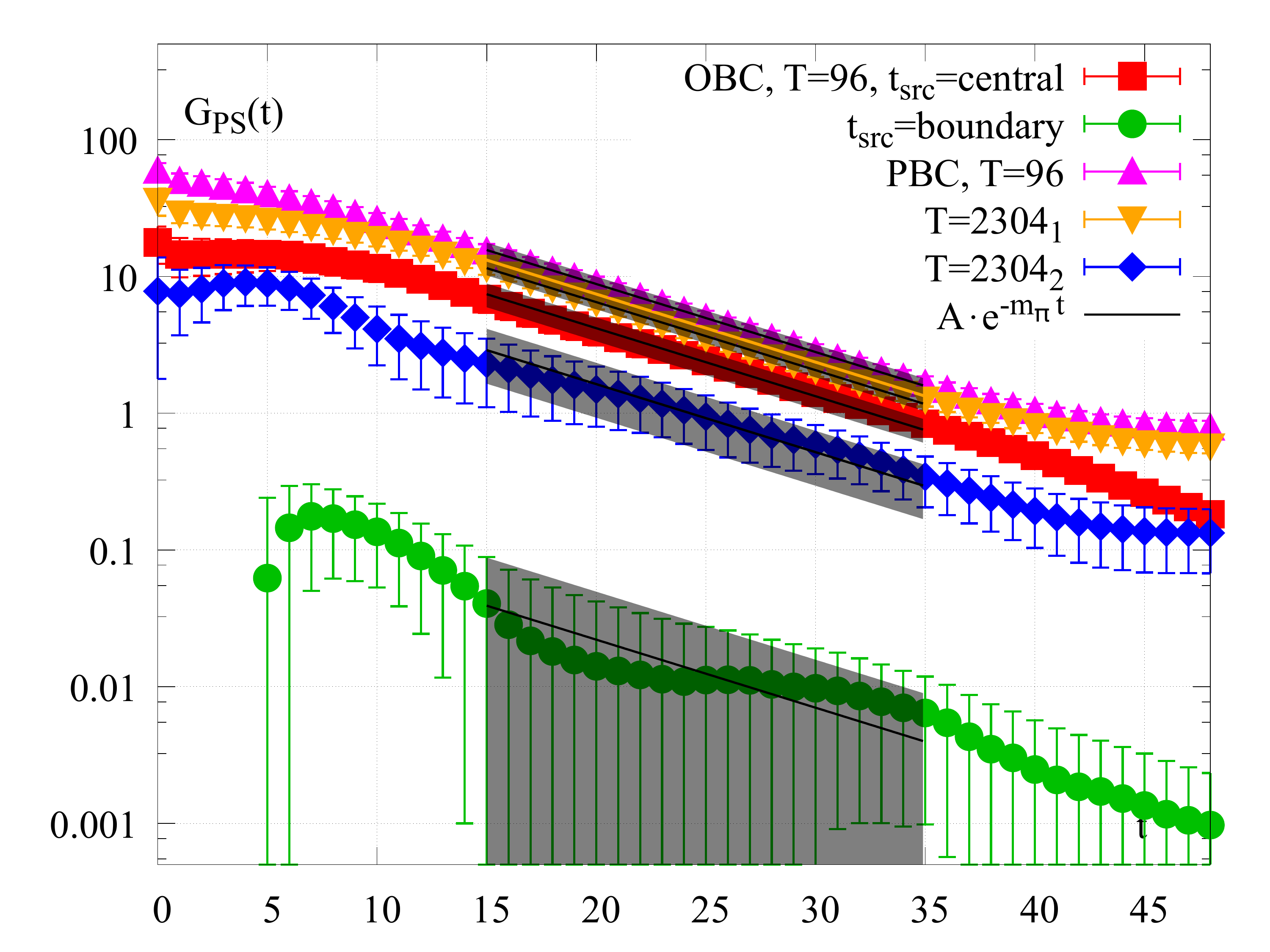}
\includegraphics[width=0.49\textwidth]{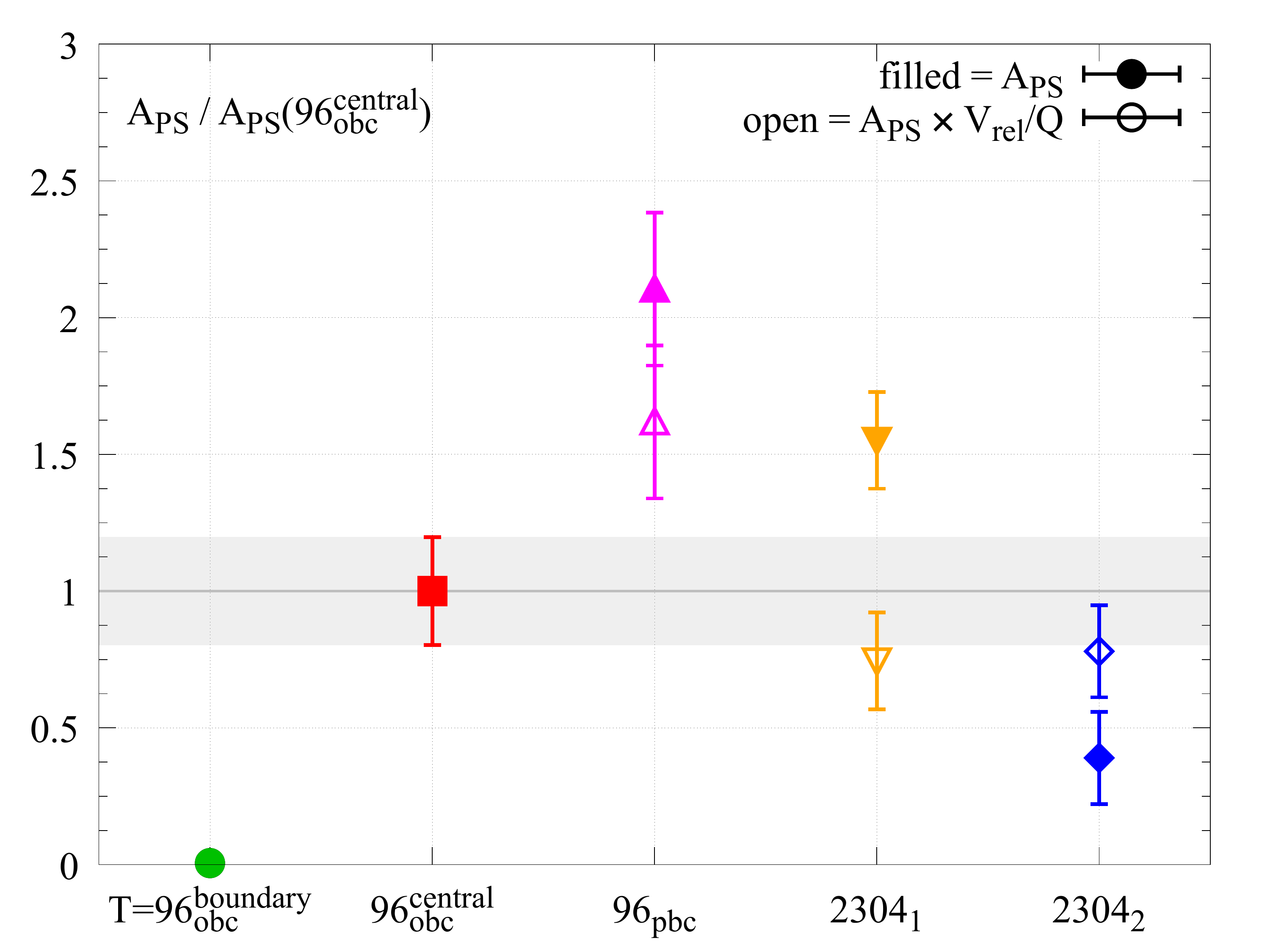}
\end{center}
\vspace{-3ex}
\caption{Left: $PS$ meson correlators on different lattice geometries. The shaded regions show single exponential fits. Right: Results for $A_{PS}$ (filled symbols) and rescaled by $V_{\rm rel}/Q$ (open symbols).}
%effective masses in the $PP$, $PS$ and $SS$ (right) for a subset of results. \alert{Non-understood short distance effect in effective mass of $SS$ correlators between P(2304)$_2$ and O(96)? Might have to drop figure.}}
\label{fig:corrs}
\end{figure}

The $PS$ correlators for the ensembles P(96), P(2304)$_1$, P(2304)$_2$ as well as O(96) with central and boundary placed sources are shown in Fig.~\ref{fig:corrs} (left). Since the sign of this correlator 
varies with $Q$, we apply an ensemble-by-ensemble sign to render it 
positive in the figure. The errors are computed using a binning procedure.
We observe that compared to the boundary placed sources the centrally placed results, and all periodic ensembles, show a larger amplitude by orders of magnitude.
%Interchanging the source and sink operators in the OBC case affects these observations. However, here we have chosen to show those that show the smallest amplitude, swapping the operators leads to an increase by roughly a factor $\sim 3$ (central placed) or $\sim 20$ (boundary placed).

Going further the correlators are fitted to a single exponential Ansatz, the results are shown as dark shaded regions in the figure. The ratio of the extracted amplitudes over the centrally placed O(96) results $A_{PS}/A_{PS}(96^{\rm central}_{\rm obc})$ is shown in Fig.~\ref{fig:corrs} (right) as filled symbols. 
For the periodic boundaries ensembles, we see that the results on P(96) show the largest contamination. Relative to the central O(96) results they are a factor 2 larger.
Comparing the P(2304)$_1$, P(2304)$_2$ we see that, while P(2304)$_1$ is larger than the central O(96) by a factor 1.5, the P(2304)$_2$ is smaller by a factor 2.
As such, even at this preliminary stage, the $T/a=2304$ lattices can be competitive.

The ratio can be further modified by multiplying it with $V_{\rm rel}/\bar{Q}$ as a rough estimate aimed at taking into account the naive expected scaling with $Q/V$. The corresponding results for the periodic boundary ensembles are shown as open symbols in Fig.~\ref{fig:corrs} (right). We observe that the P(2304)$_1$ and P(2304)$_2$ results agree with good accuracy and the expected scaling is roughly confirmed, however, we stress once more the preliminary nature of this study.

%Going further we show the effective masses of the $PP$, $PS$ and $SS$ channels in Fig.~\ref{fig:corrs} (right).
%The $PP$ results are given in black, where the black dashed lines denotes $2m_\pi$. The $PS$ data on P(96) are shown in grey and nicely align with the pion mass at long distance. Turning to the $SS$ channel, we plot the OBC data in red (central) and green (boundary). As expected the latter shows significant boundary effects precluding its use in practice.
%In blue we show the results on the long-$T$ lattice P(2304)$_2$. Comparing P(2304)$_2$ and O(96) with central sources we observe comparable errors and similarly long usable plateaus. Note, statistics can be boosted in the former case by using more of the time direction to place sources, in the latter case is is more difficult.

\section{Discussion}

In this proof-of-concept contribution we introduced the idea of using very cold lattices, i.e.~long-$T$ ensembles, as a means to reduce topological contaminations.
In this context we motivated the scalar-scalar and pseudoscalar-scalar meson correlators as sensitive probes.
We demonstrated that the $PS$ correlator is visibly affected by topological effects, even when the problem of global topological charge freezing has been addressed through open boundary conditions. Simultaneously the long-$T$ approach is shown to suppress topological contamination as well and is seen to lead to competitive results in comparison: the hadronic observables are "defrosted".

%The preliminary results derived from lattices with periodic boundary conditions up to 24 times the extent of traditional volumes indicate that contamination can be controlled to the same level as and even bettered than a traditional volume using open boundaries.

The long-$T$ approach has a number of benefits: For example, the traditional spatial volume enables the use of current spectroscopic methods with little change. Around-the-world effects are de-facto eliminated and multiple time-separated sources can be inverted at the same time, keeping the inversion cost per sample potentially as low as that of the (scaled) traditional volume.
Furthermore, the calculation of errors can benefit from a combination of traditional and translational averaging methods, thereby making the best of all worlds. Here we observed that our ensembles with $V_{\rm rel}=24$ already exhibit a stable translational error.

At the current stage, we did not observe signs of thermalisation effects. In the future a quantitative study is planned however. Furthermore we will analyse the $SS$ correlator by studying its spectrum and expected asymptotic behavior.

\section*{Acknowledgements}
This work was granted access to the HPC resources of Occigen (CINES), Jean-Zay (IDRIS) and Ir\`ene-Joliot-Curie (TGCC) under projects (2021,2022)-A0080511504 and (2021,2022)-A0080502271 by GENCI.
AF acknowledges support by the Ministry of Science and Technology Taiwan (MOST) under grant 111-2112-M-A49-018-MY2.
MTH is supported by UKRI Future Leader Fellowship MR/T019956/1 and in part by UK STFC grant ST/P000630/1. JRG acknowledges support from the Simons Foundation 
through the Simons Bridge for Postdoctoral Fellowships scheme.

%%%%%%%%%%%%%%%%%%%%%%%%%%%%%%%%%%%%%%%%%%%%%%%%%%%%%%%%

\bibliographystyle{JHEP}
\bibliography{references}

%\begin{thebibliography}{99}
%\bibitem{...}
%....
%\end{thebibliography}

\end{document}